\documentclass[prb,twocolumn]{revtex4}
\usepackage{txfonts,graphicx,bm,amssymb,color,ulem,here}

\begin{document}
	
\title{Broad Linewidth of Antiferromagnetic Spin Wave due to Electron 
Correlation}

\author{Michiyasu Mori}
\affiliation{Advanced Science Research Center, Japan Atomic Energy Agency, 
Tokai, Ibaraki 117-1195, Japan} 

\begin{abstract}
	We study magnetic excitations in a bilayer of an antiferromagnetic (AF) 
	insulator and a correlated metal, in which double occupancy is 
	forbidden. The effective action of the AF spin wave in the AF insulator is 
	derived by using the path integral formula within the second order of 
	interplane coupling. 
	The electron correlation in the correlated metal is treated by the 
	Gutzwiller approximation, which renormalizes the hopping integrals by 
	$g_t$ as proportional to the hole density. 
	The linewidth of the AF spin wave excitations originates from 
	particle-hole excitations in the correlated metal. 
	By increasing the correlation effect, i.e., by decreasing $g_t$, it is 
	found that the linewidth at low energies increases inversely proportional 
	to $g_t$.  
	The present results will also be useful for bilayers of a metal and 
	ferrimagnet.
\end{abstract}

\maketitle

\section{Introduction}
A high-$T_c$ cuprate shows superconductivity by doping carriers in a Mott 
insulator, which is an antiferromagnetic (AF) 
insulator~\cite{lee,ogata08,keimer}. 
Its magnetic excitation has been  
observed by inelastic neutron scattering (INS) 
measurement and is well described by AF 
spin waves~\cite{vaknin,yamada87,birgeneau,hayden,coldea,headings}. 
Since a cuprate is magnetically almost isotropic, the AF spin wave is 
approximately gapless, appearing at the commensurate wavenumber.  
By doping holes, on the other hand, low-energy 
magnetic 
excitations appear at incommensurate 
wavenumbers~\cite{yoshizawa,shirane89,yamada}, and the overall feature of 
magnetic excitations changes from the usual AF 
spin wave to an hourglass-like 
spectrum~\cite{tranquada04,fujita,vignolle,reznik,lipscombe,norman,yamase06}. 
This variation of magnetic excitations is apparently caused by doped holes.  
However, some experimental studies using resonant inelastic X-ray 
scattering~\cite{letacon} and INS~\cite{matsuura,sato} report that a 
high-energy part of magnetic excitation is not affected by doped holes. 
Naively thinking, both low- and high-energy parts could be changed by doping, 
while the situation is rather close to the low-energy part coming from 
a Fermi surface, i.e., metallic nature, and the high-energy part having a 
localized nature.   
Furthermore, it is also claimed that two components, i.e., commensurate and 
incommensurate, comprise the magnetic excitations at low energies~\cite{sato}. 

The recent development of neutron facilities enables us to observe magnetic 
excitations more accurately in wider regions of energy and momentum. 
In addition to the position and intensity of magnetic excitations, the 
width of peaks is also an important tool 
to extract relevant interactions of magnons and to elucidate the criticality of 
magnetism~\cite{bayrakci,bayrakci2,tseng}. 
In three dimensions, the linewidth of AF spin waves is 
proportional to $k^2$ with wavenumber $k$~\cite{harris}, which coincides with a 
prediction by hydrodynamics~\cite{halperin},  
while it becomes proportional to $k$ in two dimensions~\cite{tyc}. 
The former case of proportionality with $k^2$ originates from short-wavelength 
magnon interactions, 
whereas long-wavelength magnon interactions are relevant to the latter 
one~\cite{tyc}.  
For reference, we also consider a ferromagnet. 
Above the Curie temperature, the linewidth of 
paramagnons is proportional to $k$. 
However, near the critical temperature in some materials, e.g., UGe$_2$ and 
UCoGe, 
their linewidth is independent of $k$~\cite{chubukov}. 
The constant linewidth in a ferromagnet cannot be satisfied by one type of 
excitation. Thus, these systems require the presence of both itinerant and 
localized components~\cite{chubukov}. 
Although this situation is different from that of cuprates, it implies that the 
essential 
nature of magnetic excitations can be clarified by the 
linewidth of magnetic excitations. 

In this paper, we study a bilayer system of an AF insulator and a correlated 
metal (CM), in which double occupancy is forbidden. 
The AF spin wave in the AF insulator is calculated and its linewidth is 
estimated by second-order perturbation theory with respect to an 
interplane 
magnetic exchange interaction. The correlation effect is treated by the 
Gutzwiller approximation. 
Such a bilayer system is realized in multilayered cuprates having several 
CuO$_2$ 
planes in a unit cell. Due to the different chemical environment 
between the outer and inner planes, the carrier concentration in the outer 
plane is different from that in the inner 
one~\cite{kondo89,stasio90}. In some 
cases, an alternating stack of superconducting and AF planes is realized, as 
observed by nuclear magnetic resonance~\cite{trokiner91,mukuda12}, and has been 
studied theoretically~\cite{mori02,mori05,mori06}. 
Furthermore, a bilayer of a magnet and metal is one of the basic device 
structures in 
spintronics. For example, the spin Seebeck effect produces electricity by 
applying a temperature gradient to a bilayer of a ferromagnet and 
metal~\cite{uchida08,adachi13}. 
	In such a study, a ferrimagnet is often used as a ferromagnet. 
	However, a ferrimagnet has 
	several particular aspects different from a ferromagnet, e.g., acoustic 
	and 
	optical spin wave excitations, compensation temperature, and so 
	on~\cite{ohnuma,kikkawa,geprags}. The linewidth of a spin wave in a 
	ferrimagnet 
	is 
	crucial to improve the efficiency of the spin 
	Seebeck effect. Our results are applicable to ferrimagnetic 
	spin waves.

The rest of the paper is organized as follows. In Sect.~\ref{model}, the 
Hamiltonian of the bilayer will be given and the effective action of 
Holstein--Primakoff bosons will be derived by using the path integral formula 
and the Gutzwiller approximation. 
In Sect.~\ref{results}, the AF spin wave with the linewidth 
originating from the CM will be shown numerically. 
Finally, a summary and discussion will be given in Sect.~\ref{summary}. 
The appendix on ferrimagnets will be useful 
from the viewpoint of spintronics.

\section{Model and Method}\label{model}
\begin{figure}[t]
	\begin{center}
	\includegraphics[width=0.45\textwidth]{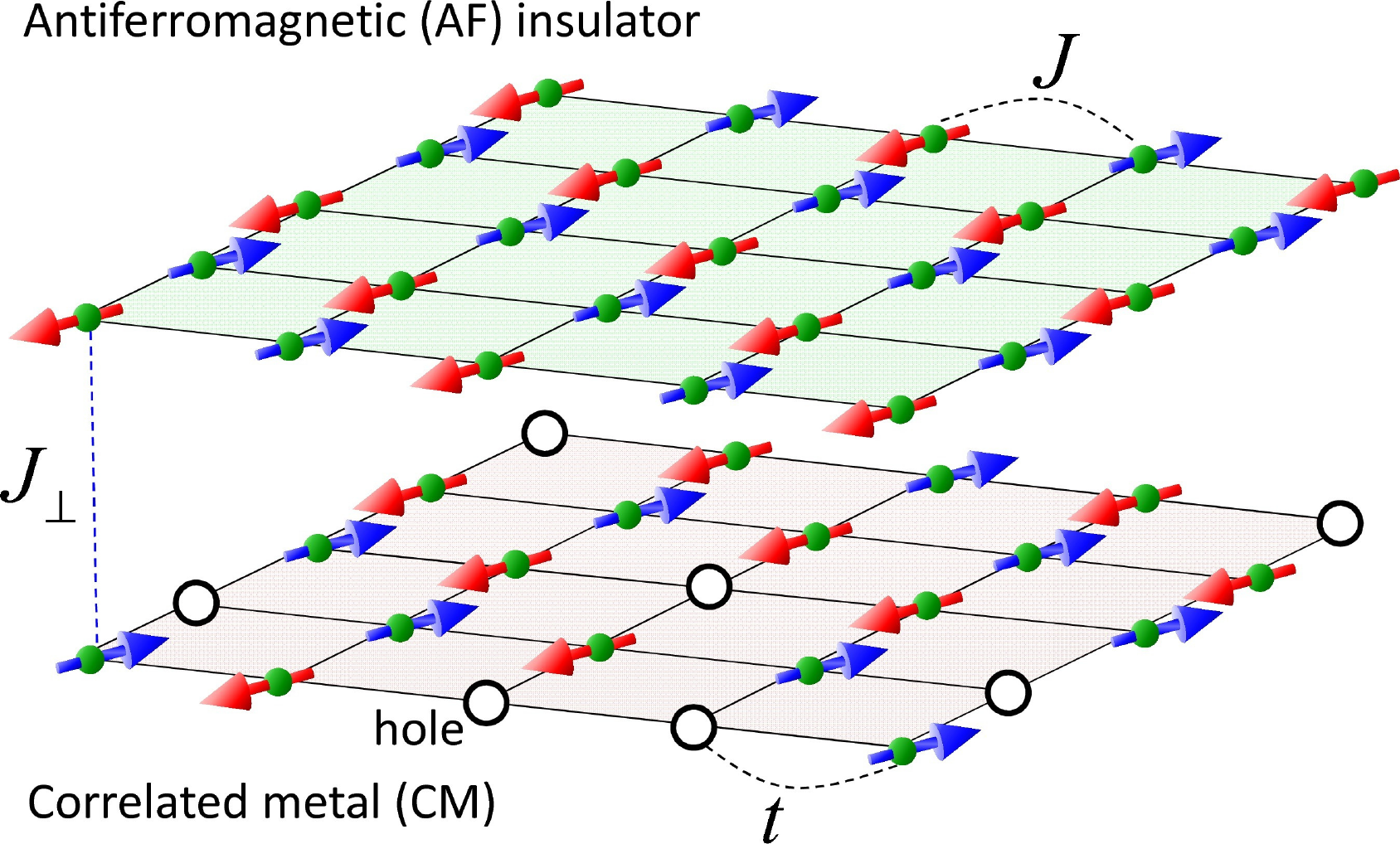}
	\caption{(Color online) Bilayer model of the AF insulator and the CM. The 
	upper plane is 
	the AF insulator and the 
	lower one is the CM, in which double occupancy is 
	forbidden. The arrows 
	indicate localized spins and the white circles indicate doped holes. The 
	magnitudes of in-plane and interplane magnetic exchange interactions are 
	denoted by $J$ 
	and 
	$J_\perp$, respectively. In this paper, $J$ in the CM is not considered. 
	The hopping integral of electrons in the CM plane is denoted by $t$ ($t'$ 
	and $t''$ are not shown in 
	this figure). Note that one-particle hopping 
	is not allowed 
	between planes due to the constraint of no double occupancy~\cite{mori06}. }
	\label{bilayer}
	\end{center}
\end{figure}

We consider the bilayer model of the AF insulator
and the CM shown in Fig. \ref{bilayer}. 
The Hamiltonian of the bilayer $H_{\rm BL}$, Eq. (\ref{bl}), is composed of 
three parts: 
$H_{\rm AF}$ for the AF insulator, $H_{\rm CM}$ for the CM, and $H_{\rm I}$ for 
the interplane 
coupling. 
The magnetic exchange interaction $J$ in Eq.~(\ref{afi}) is imposed on 
nearest-neighbor spins $\vec{S}_i$ and $\vec{S}_j$ on sites $i$ and $j$ in the 
AF 
plane, respectively.  
	Below, the lattice is divided into sublattices $A$ and $B$, and indices $i$ 
	and $j$ are associated with sublattices $A$ and $B$, 
	respectively, i.e., $i\in A$ and $j\in B$. 
	In Eq.~(\ref{metal}), electron creation (annihilation) operators with spin 
	$\sigma$ on sites $i\in 
	A$ and $j\in B$ are denoted by $c_{i,\sigma}^\dag$ ($c_{i,\sigma}$) and 
	$d_{j,\sigma}^\dag$ ($d_{j,\sigma}^{}$), respectively. 
	Each operator has a constraint of no double occupancy on each site. 
	The hopping integrals between first ($t$), second ($t'$), and third ($t''$) 
	neighboring sites are included, and $i'$ ($j'$) and $i''$ ($j''$) denote 
	first- and second-neighbor sites belonging to the same $A$ ($B$) 
	sublattice, 
	respectively. 
The interplane coupling $J_\perp$ in Eq.~(\ref{inter}) is a magnetic exchange 
interaction between the AF and CM planes. Electron hopping is forbidden 
between these planes since the AF plane does not have holes, and double 
occupancy is also forbidden \cite{mori06}. 
The spin operator on site $i$ in the CM plane is denoted by $\vec {\sigma}_i$. 
	\begin{eqnarray}
	H_{\rm BL} &=& {H_{\rm AF}} + {H_{\rm CM}} + {H_{\rm I}}\label{bl}\\
	{H_{\rm AF}} &=& J\sum\limits_{\left\langle {i,j} \right\rangle } {{{\vec 
	S}_i} 
		\cdot {{\vec S}_j}}\label{afi}\\
	{H_{\rm CM}} &=& 
	-t\sum_{\left\langle i,j \right\rangle,\sigma} 
	\left(c_{i\sigma }^\dag d_{j\sigma }^{} + h.c.\right)\nonumber\\
	&+&t'\sum_{\left\langle i,i' \right\rangle,\sigma}
	\left(c_{i\sigma }^\dag c_{i'\sigma }^{} + h.c. \right)
	-t''\sum_{\left\langle i,i'' \right\rangle,\sigma} 
	\left(c_{i\sigma }^\dag c_{i''\sigma }^{} + h.c. \right) \nonumber\\
	&+&t'\sum_{\left\langle j,j' \right\rangle,\sigma}
	\left(d_{j\sigma }^\dag d_{j'\sigma }^{} + h.c. \right)
	-t''\sum_{\left\langle j,j'' \right\rangle,\sigma} 
	\left(d_{j\sigma }^\dag d_{j''\sigma }^{} + h.c. \right) \nonumber\\
	&+& {J_ \bot }S\left( { \sum\limits_{i} {\sigma _i^z}  - 
		\sum\limits_{j} {\sigma _j^z} } \right),\label{metal}\\
	{H_{\rm I}} &=& {J_ \bot }\sum\limits_i {{{\vec S}_i}}  \cdot {{\vec \sigma 
		}_i}.\label{inter}
	\end{eqnarray}
Here, $t$=2.5, $t'$=0.85, $t''$=0.58, $J$=1, and $J_\perp$=0.1. 

	The Holstein--Primakoff bosons $a_i^\dag, a_i$ on sublattice $A$ and 
	$b_j^\dagger, b_j$ on sublattice $B$ are defined by
	$S_i^- =\sqrt{2S} a_i^\dag \left(1 - a_i^\dag a_i/(2S) \right)^{1/2}$, 
	$S_i^ + = \sqrt {2S} \left( 1 - a_i^\dag a_i/(2S)\right)^{1/2} a_i$, 
	$S_i^z = S - a_i^\dag {a_i}$, 
	$S_j^ +  = \sqrt{2S} b_j^\dag \left(1 - b_j^\dag b_j/(2S)\right)^{1/2}$, 
	$S_j^ -  = \sqrt{2S}\left(1 -b_j^\dag b_j/(2S)\right)^{1/2}b_j$, $S_j^z = 
	b_j^\dag {b_j} - S$, and $S$=1/2.
Up to the first order of the Holstein--Primakoff bosons around the N\'{e}el 
order, 
the Hamiltonian in Eq.~(\ref{bl}) is transformed as 
	\begin{eqnarray}
	{H_{\rm HP}} &=& {H_{\rm sw}} + {H_{\rm e}} + {H_\perp},\label{hp}\\
	{H_{\rm sw}} &=& JS\sum\limits_{i,\eta } {\left( { {{a_i}{b_j} + a_i^\dag 
	b_j^\dag } + a_i^\dag {a_i} + b_j^\dag {b_j}} \right)},\\
	H_{\rm e} 
	&=&H_{\rm CM} 
	+{J_ \bot }S\left( { \sum\limits_{i} {\sigma _i^z}  - 
		\sum\limits_{j} {\sigma _j^z} } \right),\label{hele}\\
	{H_\perp} &=& {J_ \bot }\sqrt{\frac{S}{2}}\left[ { \sum\limits_{i} {\left( 
			{{a_i}\sigma _i^ -  + a_i^\dag \sigma _i^ + } \right)}  + 
			\sum\limits_{j} 
		{\left( {b_j^\dag \sigma _j^ -  + {b_j}\sigma _j^ + } \right)} } 
		\right]. 
	\end{eqnarray}  

To calculate the spin wave excitation in the AF plane and the width of the 
spectrum, 
the path integral formula is useful since the magnon operators can be treated 
as 
$c$-numbers~\cite{nagaosa}. 
 The action $S$ is given by 
  \begin{widetext}
 \begin{eqnarray}
 S&=&S_0+S_1,\\
 S_0&=&\sum_{q,i\nu_n} 
	 \Phi^\dag\left[
		 \left(
		 \begin{array}{*{20}{c}}
 			-i\nu_n&0\\
 			0&i\nu_n
		 \end{array}
		 \right)
	+z_1SJ\left(
			\begin{array}{*{20}{c}}
 			1&\gamma_q^*\\
	 		\gamma_q&1
 			\end{array}
 		  \right)
 	\right]\Phi,\label{actspin}\\
 S_1 &=&-\frac{1}{2}\sum_{k,q,m,n}
	 \Psi^\dag\left[
		\left(
		\begin{array}{cc}
		i\omega_m &0\\
		0&i\omega_m+i\nu_n
		\end{array}
		\right)		
		-\left(
		\begin{array}{cc}
			H(k,i\omega_m)&\lambda M(q,i\nu_n)\\
			\lambda M^\dag(q,i\nu_n)&H(k+q,i\omega_m+i\nu_n)
		\end{array}
		\right)\right]
		\Psi, \label{action}\\
\gamma_q &\equiv& \frac{1}{z_1}\sum_{\rm e_i}e^{i{\bf q}{\bf e}_i},
\label{gamma}		
\end{eqnarray}
\end{widetext}
	where $k$ and $q$ are the momentum of electrons and magnons, respectively. 
	The 
	Matsubara frequencies of fermions and bosons are denoted by $\omega_n$ and 
	$\nu_n$, respectively. 
In Eq. (\ref{gamma}), ${\bf e}_i$ runs over the nearest-neighbor sites, and 
hence $z_1$=4 since the magnetic exchange interaction $J$ is supposed only on 
the neighboring bonds.
The field operators of electrons $\Psi$ and magnons $\Phi$ are defined as
\begin{eqnarray}
\Phi^\dag&\equiv&\left(a_q^\dagger(i\nu_n),b_{-q}(-i\nu_n)\right),\\
\Psi^\dag&\equiv&\left(\psi_k^\dag(i\omega_m),\psi_{k+q}^\dag(i\omega_m+i\nu_n)\right),\\
\psi_{k}^\dag(i\omega_m) &\equiv& \left(
c_{k \uparrow  }^\dag(i\omega_m), 
c_{k \downarrow}^\dag(i\omega_m), 
d_{k \uparrow  }^\dag(i\omega_m), 
d_{k \downarrow}^\dag(i\omega_m)
\right). 
\end{eqnarray}
In Eq.~(\ref{action}), each matrix element is given by
\begin{eqnarray}
H(k,i{\omega _n}) 
&\equiv& 
\left( {\begin{array}{*{20}{c}}
	\eta _k+h & 0 & \varepsilon_k & 0\\
	0 & \eta _k- h & 0 & \varepsilon _k\\
	\varepsilon_k & 0 & \eta_k - h & 0\\
	0 & \varepsilon_k & 0 & \eta _k + h
	\end{array}} 
\right),\label{hdiag}\\
M(q,i{\nu _n})
&\equiv&
\left( {\begin{array}{*{20}{c}}
	0&a_q^\dag (i{\nu_n})&0&0\\
	a_{-q}( - i{\nu_n})&0&0&0\\
	0&0&0&b_{-q}(-i{\nu_n})\\
	0&0&b_q^\dag (i{\nu_n})&0
	\end{array}} 
\right),\label{hoff}\\
\varepsilon_k&\equiv& -2tg_t\left[\cos(k_x)+\cos(k_y)\right],\\
\eta_k&\equiv&+4t'g_t\cos(k_x)\cos(k_y)\nonumber\\
&&-2t''g_t\left[\cos(2k_x)+\cos(2k_y)\right]
-\mu,
\end{eqnarray}
with
$k'-k=q$, 
$\lambda\equiv{J_\perp}\sqrt{S/2\beta}$, 
$h \equiv  J_\perp S/2$, and $\beta$ is the inverse of the temperature $T$. 
Below, $\beta$=10 is adopted. 
In the 
coupling between electrons and magnons, $\beta$ remains due to the definition 
of the Fourier transformation of field operators, i.e., 
$c(\tau)\equiv\sqrt{1/\beta}\sum_m \exp(-i\omega_m\tau)c_m$ with imaginary time 
$\tau$. 
The Gutzwiller approximation is adopted in the kinetic terms as 
$g_t$=$2p/(1+p)$, where $p$ is the hole concentration. 
Integrating out the electron degree of freedom within the second order of 
$J_\perp$ leads to the self-energy of the spin wave as 
	$\Sigma = - (1/2){\rm Tr} 
	\left[\hat{g}(k)M(q,i\nu_n)\hat{g}(k+q)M^\dag(q,\nu_n)\right]$ 
with the electron Green function  
	$\hat{g}(k)\equiv [i\omega_m - H(k,i\omega_m)]^{-1}$. 
After integration with respect to the Matsubara frequency of electrons 
$i\omega_m$, we can obtain the effective action of magnons $S_{\rm eff}$ as 
\begin{widetext}
\begin{eqnarray}
S_{\rm eff} 
	&=& S_0 + \Sigma,\\
\Sigma
	&=& J_\bot^2S\sum_{q,i\nu_n}
	\Phi^\dag
	\left(
	\begin{array}{*{20}{c}}
	A_+&B\\
	B  &A_-
	\end{array}
	\right)\Phi, \label{sigma}\\
A_\pm&=& \frac{1}{\beta}\sum_{i\omega_m,k}
		\frac{(i\omega_m+i\nu_n-\eta_{k+q}\mp h)
			  \left(i\omega_m-\eta_k\pm h\right)}
			{\left[(i\omega_m+i\nu_n-\eta_{k+q})^2-E_{k+q}^2\right]
			 \left[\left(i\omega_m-\eta_k\right)^2-E_k^2\right]},\label{p-h-A}\\
B	&=& \frac{1}{\beta}\sum_{i\omega_m,k}
\frac{\varepsilon_{k+q}\varepsilon_k}
{\left[(i\omega_m+i\nu_n-\eta_{k+q})^2-E_{k+q}^2\right]
	\left[\left(i\omega_m-\eta_k\right)^2-E_k^2\right]},\label{p-h-B}
\end{eqnarray}
\end{widetext}
with $E_k=\sqrt{\varepsilon_k^2+h^2}$. 
Hence, the Green functions of magnons $\hat{G}(q,i\nu)$ are given by  
\begin{eqnarray}
\hat{G}(q,i\nu_n)
	&\equiv& 
	\left\langle 
	\left(
	\begin{array}{*{20}{c}}
	a_q(i{\nu_n})a_q^\dag (i{\nu_n})&a_q(i{\nu_n})b_{-q}(-i{\nu_n})\\
	b_{-q}^\dag(-i{\nu_n})a_q^\dag (i{\nu_n})&b_{-q}^\dag(-i{\nu_n})b_{-q}(-i{\nu_n})
	\end{array}	
	\right)
	\right\rangle,\nonumber\\
	&=&
	\left(
	\left(
	\begin{array}{*{20}{c}}
	-i\nu_n&0\\
	0&i\nu_n
	\end{array}	
	\right)
	+z_1JS
	\left(
	\begin{array}{*{20}{c}}
	1+\alpha A_+&\gamma_q+\alpha B\\
	\gamma_q+\alpha B&1+\alpha A_-
	\end{array}	
	\right)
	\right)^{-1}\label{green}
\end{eqnarray}
with $\alpha  \equiv J_\bot^2/(Jz_1)$.
The dispersion relation of the spin wave is obtained as
\begin{widetext}
\begin{eqnarray}
E_{\pm}(\nu)  
&=&
	 {z_1}SJ\left\{\pm\frac{\alpha }{2}\left(A_+ - A_- \right) 
	 	+ \sqrt{
		 	\left(1-\gamma_q^2\right)
		 	+2\alpha\left(\frac{A_+ + A_-}{2}-B\gamma_q \right)
		 	+\alpha^2
			 	\left[
			 	\left(\frac{A_+ + A_-}{2}\right)^2-B^2
			 	\right]
		 	} \right\}. \label{mode}
\end{eqnarray}
\end{widetext}
Below, $\nu$ is interpreted as $\nu+i\delta$ with constant $\delta$=0.1 and 
$\nu>$0. 
Note that the ferrimagnetic case is given in the Appendix. 
In the first term of Eq.~(\ref{mode}), the negative sign is chosen to satisfy 
causality and to obtain the well-defined spectrum. 

\section{Results}\label{results}
\begin{figure}[h]
	\begin{center}
		\includegraphics[width=0.43\textwidth]{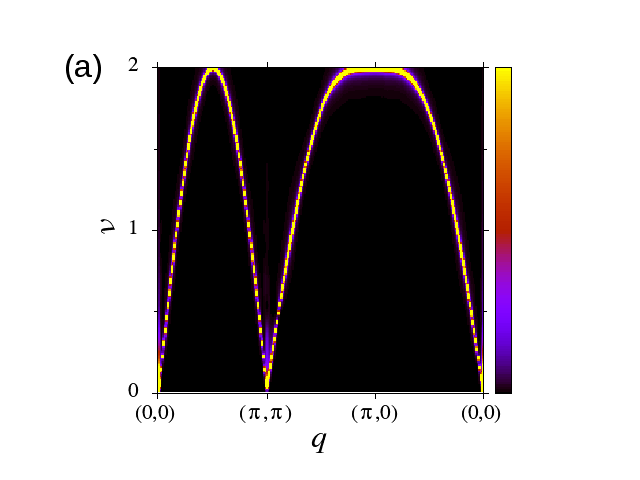}
		\includegraphics[width=0.43\textwidth]{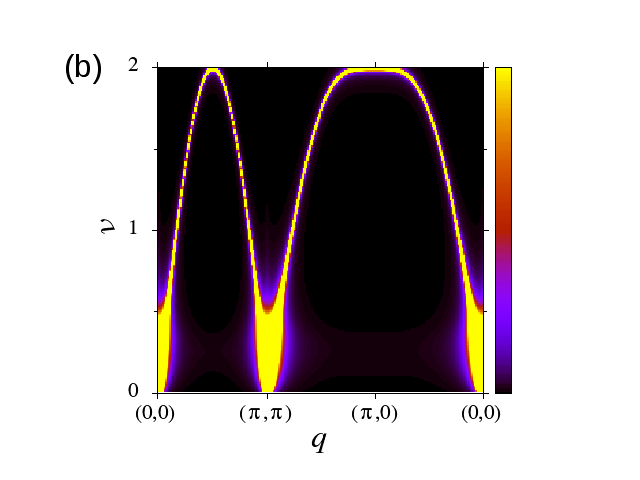}
		\includegraphics[width=0.4\textwidth]{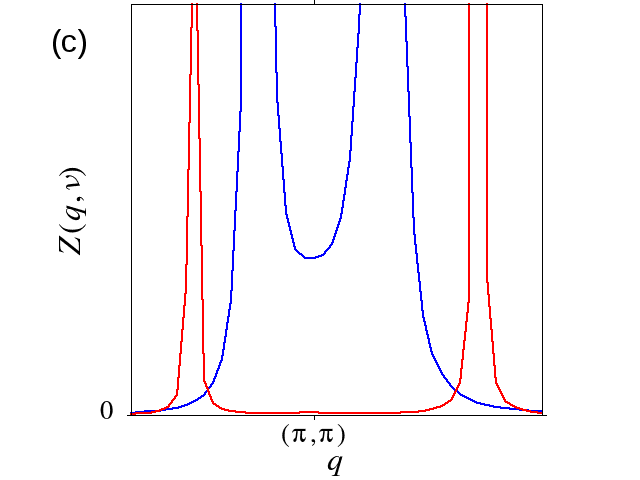}
		\caption{(Color online) Spin wave excitation in the AF plane for (a) 
			non-correlated case, 
			$g_t=1$, and (b) correlated case, $g_t$=$2p/(1+p)$ with $p$=0.01. 
			(c) 
			$\nu$-cuts of $Z(q,\nu)$ around ($\pi$,$\pi$) plotted for $\nu$=0.5 
			(blue solid 
			line) and 1.0 (red solid line).}  
		\label{f2}
	\end{center}
\end{figure}

To see the dispersion relation of the AF spin wave, we plot the spectral 
function 
defined by  
\begin{equation}
Z(q,\nu)
	\equiv
	-\frac{1}{\pi}{\rm Im}\left[\frac{1}{\nu-E_-(\nu)}\right]^{-1},
\end{equation}
for (i) the non-correlated case, $g_t=1$, and (ii) the correlated case, 
$g_t$=$2p/(1+p)$ with $p$=0.01, in Figs. \ref{f2}(a) and 2(b), respectively. In 
Fig.~\ref{f2}(a), one can see the usual dispersion relation of the AF spin 
wave 
on the square lattice. The linewidth almost does not changes. 
In the correlated case, on the other hand, it is clear that the low-energy 
region of the spectrum is broad as shown in Fig.~\ref{f2}(b). 
To see the broadening of the spectrum, their intensities are plotted in Fig. 
\ref{f2}(c) for $\nu$=0.5 (blue solid line) and $\nu$=1.0 (red solid line). 
The broadening of the spectrum is marked in the low-energy region, i.e., 
$\nu\lesssim$ 0.5.

\begin{figure}[h]
	\begin{center}
		\hspace{-0.5cm}\includegraphics[width=0.43\textwidth]{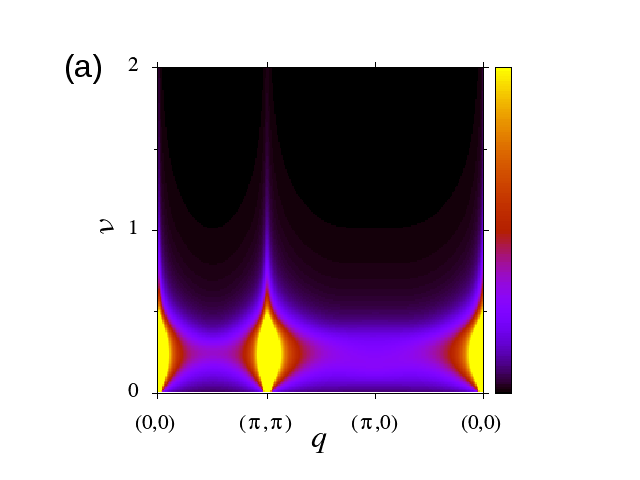}
		\includegraphics[width=0.38\textwidth]{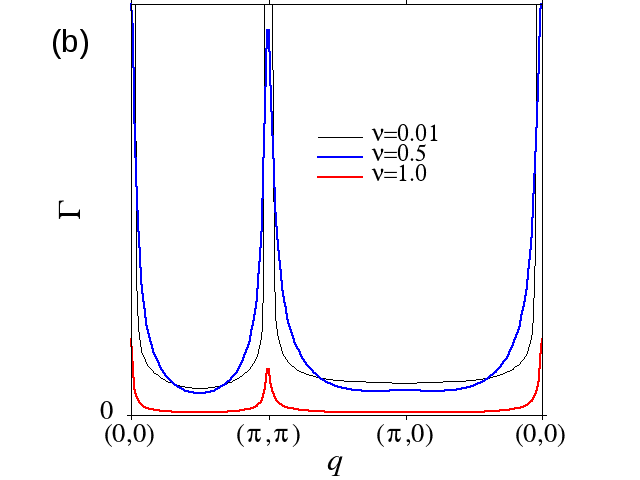}
		\includegraphics[width=0.38\textwidth]{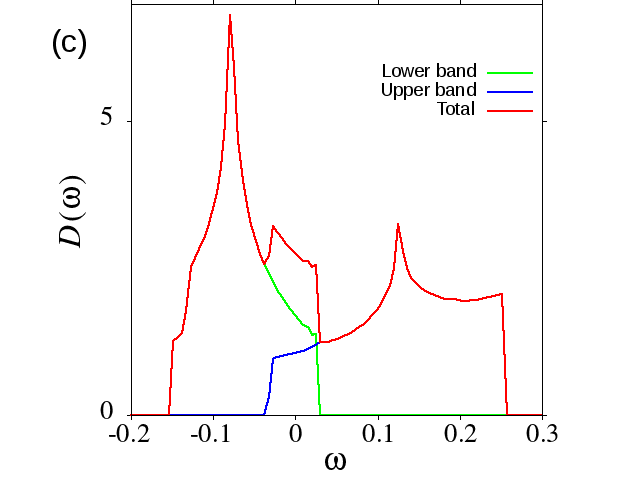}
		\caption{(Color online) (a) Imaginary part of $E_-(q,\nu)$, which 
			corresponds to the 
			width $\Gamma$, plotted in ($q$,$\nu$) plane for the correlated 
			case 
			with $p$=0.01. (b) $\nu$-cuts of Im$E_-(q,\nu)$ for $\nu$=0.01 
			(gray), 
			0.5 (blue), and 1.0 (red). (c) Density of states (DOS) in the CM 
			with 
			$p$=0.01. 
			The dispersion relation of 
			electrons in the CM is folded by the AF order. The DOS of the upper 
			(lower) band is plotted by the blue (green) line. The total DOS, 
			i.e., 
			the sum of the upper and lower bands, is shown by the red line. The 
			DOS 
			of the non-correlated case is 
			not 
			shown here since its magnitude is two orders smaller than those 
			shown 
			in 
			this figure.}
		\label{f3}
	\end{center}
\end{figure}

The imaginary part of $E_-(q,\nu)$, which corresponds to the width of spectrum 
$\Gamma$, is plotted in Fig. \ref{f3}. Figure \ref{f3}(a) shows the 
distribution 
of ${\rm Im}E_-(q,\nu)$ on the ($q$,$\nu$) plane, while several $\nu$-cuts for 
the 
correlated case with $p$=0.01 are plotted in Fig. \ref{f3}(b). 
Again, we can see that $\Gamma$ is large in the region of $\nu\lesssim$ 0.5, 
particularly around $\nu\sim$ 0.25, as shown in Fig.~\ref{f3}(a). In addition, 
the large 
$\Gamma$ is rather 
concentrated around (0,0) and ($\pi$,$\pi$) in Fig.~\ref{f3}(b).
Here, the question may arise; why is the low-energy region so broad? 
To answer this question, the density of states, $D(\omega)$, in the CM is 
plotted in Fig.~\ref{f3} (c) for $g_t$=$2p/(1+p)$ with $p$=0.01. 
The band in the CM is folded by the AF order in the AF plane, i.e., the 
last term in Eq.~(\ref{hele}), and it comprises upper and lower bands.   
The AF order is assumed to be stable in this paper.  
It is important to note the suppression of the band width $W$ to $\sim$ 0.5 due 
to the 
Gutzwiller factor $g_t\sim$0.02. 
The magnitude of $D(\omega)$ is 
automatically enhanced to conserve the total number of states. 
In fact, $D(\omega)$ for the non-correlated case is two orders smaller than 
that 
for the correlated one and is not visible in Fig.~\ref{f3} (c). 
Note that $\Sigma$ is composed of particle-hole 
excitations in the CM, which can be seen from Eqs.~(\ref{p-h-A}) and 
(\ref{p-h-B}), which leads to $\Gamma$.
When $W$ is suppressed and the magnitude of $D(\omega)$ is 
enhanced, the particle-hole excitations gain a large phase volume at low 
energies, i.e., $\omega\lesssim$ 0.5. 
In particular, the particle-hole excitations between the two peaks 
corresponding to 
the van Hove singularity in Fig.~\ref{f3} (c) have the largest contribution, 
whose energy is about $\omega\sim$ 0.25.  
In fact, $\Gamma$ is large around $\nu\sim$0.25 as shown in 
Fig.~\ref{f3}~(a). 
In the non-correlated case with the same energy window $\sim$0.5, the phase 
volume of 
particle-hole excitations is two orders smaller than that in the correlated 
case since its band width is two orders larger than that in the correlated 
case. 
This is the reason why we could not see any broadening in Fig.~\ref{f2}(a). 
On the whole, $\Gamma$ is large around (0,0) and ($\pi$,$\pi$) as shown 
in Fig.~\ref{f3}(b).
Since the origin of $\Gamma$ is particle-hole excitations in the CM, this 
enhancement implies the nesting of the Fermi surface near half-filling with 
the band folding.

\begin{figure}[h]
	 	\begin{center}
	 		\includegraphics[width=0.44\textwidth]{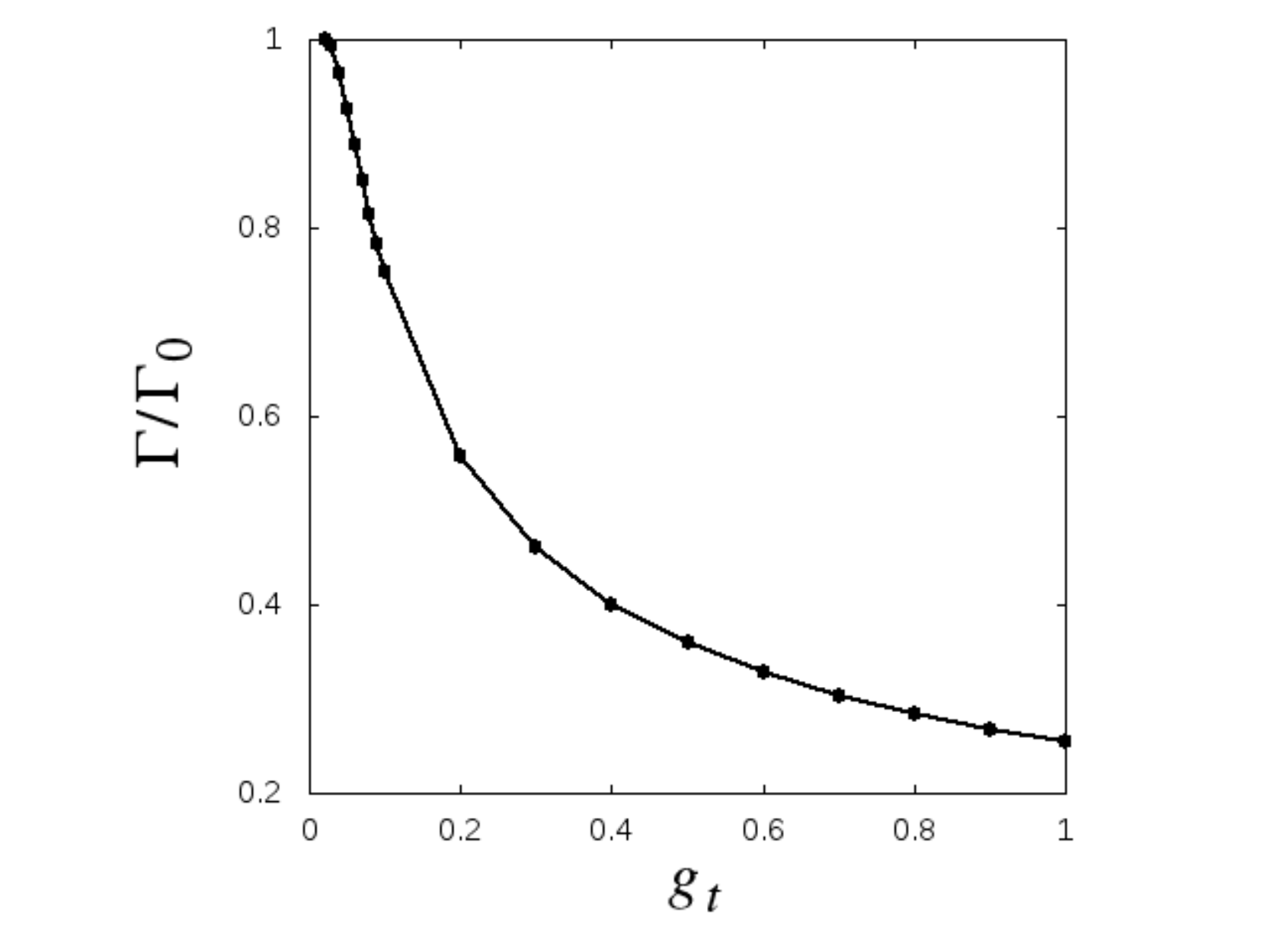}
	 		\caption{
	 			$\Gamma$ plotted as a function of $g_t$ for $q$=($\pi$,$\pi$), 
	 			$\nu$=0.5. It is normalized by its magnitude at 
	 			$g_t$=0.02, defined by $\Gamma_0$. Note that $p$ is fixed to 
	 			0.01 
	 			and 
	 			$g_t$ is exceptionally assumed to be independent of $p$ in 
	 			order to 
	 			simulate the correlation effect on 
	 			$\Gamma$. } 
	 		\label{gtdep}
	 	\end{center}
	 \end{figure}

	So far, we have compared the non-correlated case, $g_t$=1, with only 
	one correlated case, $g_t\sim$0.02 with $p$=0.01. 
	A quantitative estimation of $\Gamma$ is necessary. 
	In our results, the enhancement of $\Gamma$ originates 
	from $g_t$ as a function of $p$, while $\Gamma$ depends on not only $g_t$ 
	but 
	also $q$ 
	and $\nu$. 
	In Fig.~\ref{gtdep}, $\Gamma$ is plotted as a function of $g_t$ 
	 for $q$=($\pi$,$\pi$), $\nu$=0.5, and $p$=0.01. 
	It is normalized by its magnitude at $g_t$=0.02,  
	defined by $\Gamma_0$, i.e., the peak height of the blue line at 
	($\pi$,$\pi$) 
	in Fig.~\ref{f3}(b). 
	Note that $g_t$ in Fig.~\ref{gtdep} is exceptionally assumed to 
	be 
	independent of $p$. 
	Figure~\ref{gtdep} shows how $\Gamma$ is quantitatively enhanced by the 
	correlation, i.e., $g_t$. 
	It is found that $\Gamma$ increases with decreasing $g_t$. 
	Roughly estimated, it is inversely proportional to $g_t$ since $\Gamma$ 
	originates from the particle-hole excitations in the CM with the bandwidth 
	suppressed 
	by $g_t$.

\begin{figure}[h]
	\begin{center}
		\includegraphics[width=0.45\textwidth]{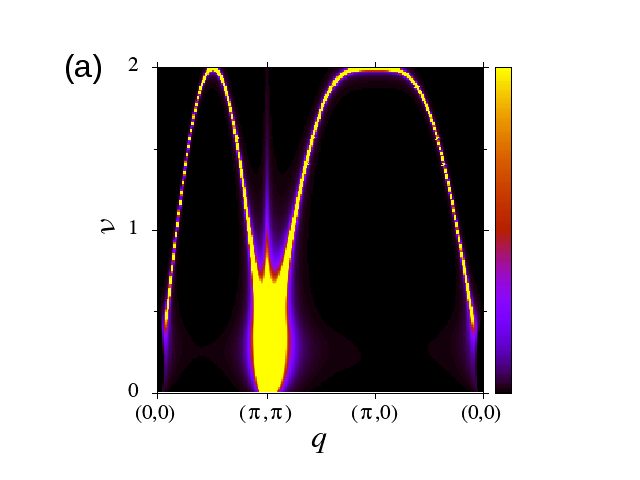}
		\includegraphics[width=0.4\textwidth]{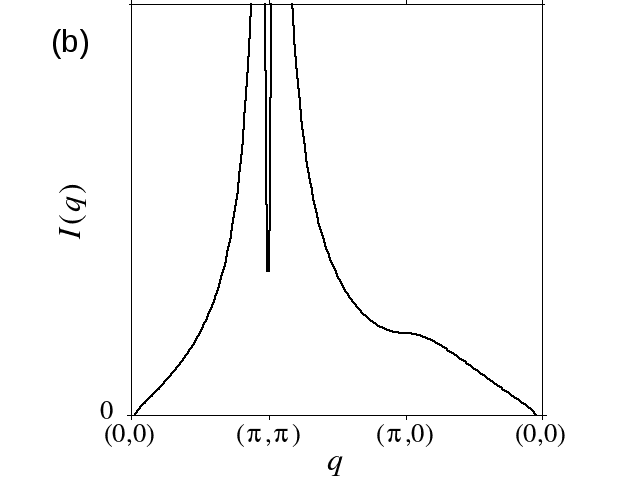}\label{ins}
		\caption{(Color online) (a) Imaginary part of spin-spin correlation 
			function, 
			$\chi^{+-}(q,\omega)$, which 
			is observed by INS measurement, for the 
			correlated case with $p$=0.01. (b) $\nu$-integrated intensity of 
			$\chi^{+-}(q,\omega)$. The intensity is enhanced around 
			($\pi$,$\pi$), 
			whereas it is suppressed around (0,0). }
		\label{f4}
	\end{center}
\end{figure}

The magnetic excitation spectrum can be measured by INS measurement, in 
which the spectrum is determined by the following correlation function: 
\begin{eqnarray}
\chi^{+-}(q,\nu)
	&=& 
	\sum_l\sum_{\alpha,\beta=A,B}
	\int \frac{dt}{2\pi}
	 e^{-i\nu t+iq{R_l}}\left\langle 
	{S_{0\alpha}^ + (0)S_{l\beta}^ - (t)} \right\rangle,\\
	&=&
	\frac{S}{\pi}\sum_{i,j=1,2}G_{i,j}(q,\nu).
\end{eqnarray}
Each component of the Green function in Eq.~(\ref{green}) is denoted by 
$G_{i,j}(q,\nu)$.  
In Fig. \ref{f4}, Im~$\chi^{+-}(q,\nu)$ is plotted for the correlated case 
$g_t=2p/(1+p)$ with $p$=0.01. 
Compared with Fig. \ref{f2}(b), the spectral weight is enhanced around 
($\pi$,$\pi$), while it is suppressed around (0,0). 
The $\nu$-integrated intensity, defined by $I(q)\equiv \int d\nu~{\rm 
Im}~\chi^{+-}(q,\nu)$, is shown in Fig. \ref{f4}(b). 
This is characteristic to the AF spin wave and is observed in 
cuprates~\cite{headings}. 
Since the INS measurement can observe the magnetic excitations around 
($\pi$,$\pi$), the broad linewidth of the AF spin wave due to the electron 
correlation can be measured by INS.

\section{Summary and Discussion}\label{summary}
We have studied the magnetic excitations in the bilayer of an antiferromagnetic 
(AF) insulator and a correlated metal (CM), i.e., a doped Mott insulator, in 
which double occupancy is forbidden. 
The correlation effect in the metallic plane is treated by the Gutzwiller 
approximation, which renormalizes the hopping integrals proportional to the 
hole 
density. The effective action of a linearized AF spin wave in the AF insulator 
is 
calculated by second-order perturbation theory with respect to the 
interplane coupling. 
The path integral formula is useful to integrate out the electron degrees of 
freedom in the CM. 
Near half-filling, the strong repulsion between electrons suppresses their 
kinetic energy and results in a narrow band width, which gives enough phase 
space for low-energy magnetic excitations. The self-energy of an AF spin wave 
originates from the particle-hole excitation in the metallic plane. 
Hence, these excitations with a narrow band width make the AF spin wave 
broad at low energies due to the electron correlation.    
	By increasing the correlation effect, i.e., by decreasing $g_t$, the 
	linewidth at low energies increases inversely proportional to 
	$g_t$.

In this paper, only the AF insulating plane was focused on, while we should 
also consider magnetic excitations in the metallic plane. 
Preceding theoretical studies showed that magnetic excitations will appear 
around incommensurate wave numbers related to the Fermi surface 
geometry~\cite{yamase06,norman}. 
In addition, the magnetic exchange interaction in the CM induces 
other phases 
such as antiferromagnetism, $d$-wave superconductivity (dSC), the flux phase, 
the singlet resonant valence bond (RVB) state, and so on. 
These states will also make the AF spin wave broad, while it will be different 
from the present case. 
In some cases, the spectral weight of an AF spin wave in the AF insulating 
plane 
will be reduced by solving magnetic excitations in both planes 
self-consistently. 
This might lead to the two components proposed in INS studies~\cite{sato}, which
will be clarified in the near future. 

If the CM plane is substituted by the dSC, such a bilayer system is a 
coexistent phase of AF and dSC, as observed in the multilayered cuprates by 
nuclear magnetic resonance (NMR)  
measurements~\cite{mukuda12}. 
Simultaneously, the coexistence within one CuO$_2$ plane has also 
been observed by NMR measurements~\cite{mukuda12} and theoretically 
studied~\cite{inaba,yamase,hayashi13}. 
In addition, the phase separation between AF and dSC states may also be 
possible\cite{koikegami}. 
Even though the magnetic excitations in these cases are not obvious, their 
situations are close to the present study. 
For example, in terms of a slave fermion with a large-$N$ 
expansion~\cite{yoshioka}, the nearest-neighbor hopping is associated with the 
interplane coupling in our bilayer 
model, and the metallic plane corresponds to the kinetic terms of the second- 
and third-neighbor hoppings. 
Hence, our results will also be useful to study the magnetic excitations in 
the coexistent and phase-separated phases. 

Finally, it is also noted that the present model is closely related to some 
spintronics devices using a bilayer of a metal and a ferrimagnet, which is 
often referred to as a ferromagnet~\cite{maekawa}.
For example, electricity can be generated by applying a temperature 
gradient to such a bilayer system. This is called the spin Seebeck effect, 
which is a 
completely new method of thermoelectric generation~\cite{uchida08,adachi13}. 
The low-energy magnetic excitations and their lifetime are crucial to 
understand the experimental signal and to improve the  
efficiency of the spin Seebeck effect~\cite{ohnuma,kikkawa}. 
The present results will contribute to such various research fields.

\begin{acknowledgments}
The author thanks S. Maekawa, M. Fujita, M. Matsuura, and S. Shamoto 
for useful discussions and helpful comments. 
This work 
was supported by Grants-in-Aid for Scientific Research (Grant No.15K05192, 
No.15K03553, and 16H01082) from JSPS and MEXT, and by the inter-university 
cooperative research program of IMR Tohoku University. Part of the numerical 
calculation was done with the supercomputer of JAEA. 
\end{acknowledgments}

\appendix
\section{Bilayer of $\bm{Ferri}$magnet and Metal}
\begin{figure}[h]
	\begin{center}
	\includegraphics[width=0.45\textwidth]{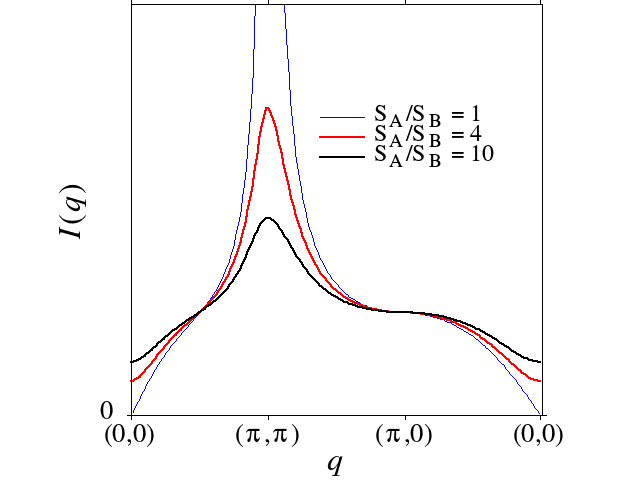}
	\caption{(Color online) $\nu$-integrated intensity of 
		$\chi^{+-}(q,\omega)$ in the case of a {\it ferri}magnet. Upon 
		increasing 
		the 
		difference between sublattice magnetizations, the enhancement around 
		($\pi$,$\pi$) is suppressed and the intensity around (0,0) increases 
		due to the opening of a gap. }
	\label{ferri}
	\end{center}
\end{figure}

We show the effective action of linearized spin waves in 
	a bilayer of a ferrimagnet and a (correlated) metal, which is often used in 
	spintronics. 
The simplest model of a ferrimagnet is two sublattices with different 
magnitudes of magnetization $S_A \neq S_B$. Instead of Eq. (\ref{afi}), the 
Hamiltonian of the ferrimagnetic plane $\tilde{H}_{\rm I}$ is given by  
\begin{equation}
\tilde{H}_{\rm I}
	= J  \sum_{\left\langle {i,j } \right\rangle } \vec{S}_i \cdot \vec{S}_j
	 -J_A\sum_{\left\langle {i,i'} \right\rangle } \vec{S}_i \cdot \vec{S}_{i'}
	 -J_B\sum_{\left\langle {j,j'} \right\rangle } \vec{S}_j \cdot \vec{S}_{j'}.
\end{equation}
Hence, the Hamiltonian in Eq.~(\ref{hp}) is substituted by  
\begin{eqnarray}
\tilde{H}_{HP} 
	&=& 
		\tilde{H}_e + \tilde{H}_{\rm int} + \tilde{H}_m\\
\tilde{H}_e  
	&=&H_{\rm CM}
	+ J_\perp\left(S_A\sum_{i\in A}\sigma_i^z - S_B\sum_{j\in B}\sigma_j^z 
	\right)\\
\tilde{H}_{\rm int}
	&=& {J_ \bot }\left[ {\sqrt {\frac{{{S_A}}}{2}} \sum\limits_{i \in A} {\left( {{a_i}\sigma _i^ -  + a_i^\dag \sigma _i^ + } \right)}  + \sqrt {\frac{{{S_B}}}{2}} \sum\limits_{j \in B} {\left( {b_j^\dag \sigma _j^ -  + {b_j}\sigma _j^ + } \right)} } \right],\\
\tilde{H}_m
	&=& J\sum_{i,j}\left\{ {\sqrt {{S_A}{S_B}} \left( {{a_i}{b_j} + a_i^\dag b_j^\dag } \right) + {S_B}a_i^\dag {a_i} + {S_A}b_j^\dag {b_j}} \right\} \nonumber\\
	&-&J_A S_A \sum_{i,i'} \left( {a_i^\dag {a_{i'}} + a_{i'}^\dag {a_i} - a_i^\dag {a_i} - a_{i'}^\dag {a_{i'}}} \right)\nonumber\\
	&-&J_B S_B \sum_{j,j'} \left( {b_j^\dag {b_{j'}} + b_{j'}^\dag {b_j} - b_j^\dag {b_j} - b_{j'}^\dag {b_{j'}}} \right).
\end{eqnarray}
Therefore, the matrix elements in Eqs. (\ref{actspin}), (\ref{hdiag}), and 
(\ref{hoff}) are replaced by
\begin{widetext}
\begin{eqnarray}
  \tilde{S}_0&=&\sum_{q,i\nu_n} 
 \Phi^\dag\left[
 \left(
 \begin{array}{*{20}{c}}
 -i\nu_n&0\\
 0&i\nu_n
 \end{array}
 \right)
 +\left(
 \begin{array}{*{20}{c}}
 z_1JS_B+z_2J_AS_A(1-\zeta_q)   & z_1J\sqrt{S_AS_B}\gamma_q\\
 z_1J\sqrt{S_AS_B}\gamma_q & z_1JS_A+z_2J_BS_B(1-\zeta_q)
 \end{array}
 \right)
 \right]\Phi,\\
\tilde{H}(k,i{\omega _n}) 
&\equiv& 
\left( {\begin{array}{*{20}{c}}
	{{\eta _k}  + {m_A}}&0&{{\varepsilon _k}}&0\\
	0&{{\eta_k}  - {m_A}}&0&{{\varepsilon _k}}\\
	{{\varepsilon _k}}&0&{{\eta _k}  - {m_B}}&0\\
	0&{{\varepsilon _k}}&0&{{\eta _k} + {m_B}}
	\end{array}} 
\right),\\
\tilde{M}(q,i{\nu _n})
&\equiv&
\left( {\begin{array}{*{20}{c}}
	0&{{\lambda _A}a_q^\dag (i{\nu _n})}&0&0\\
	{{\lambda _A}{a_q}( - i{\nu _n})}&0&0&0\\
	0&0&0&{{\lambda _B}{b_{-q}}( - i{\nu _n})}\\
	0&0&{{\lambda _B}b_{-q}^\dag (i{\nu _n})}&0
	\end{array}} 
\right),
 \end{eqnarray}
 \end{widetext}
 with 
 $\lambda_A\equiv{J_\perp}\sqrt{S_A/2\beta}$,
 $\lambda_B\equiv{J_\perp}\sqrt{S_B/2\beta}$, 
 $m_A \equiv  J_\perp S_A/2$, and 
 $m_B \equiv  J_\perp S_B/2$. 
 Hence, the self-energy, Eq.~(\ref{sigma}), is written as
 \begin{widetext}
\begin{eqnarray}
 \tilde{\Sigma}
 &=&{J_ \bot ^2 \sqrt{S_A S_B}}\sum_{q,i\nu_n}
 \Phi^\dag
 \left(
 \begin{array}{*{20}{c}}
 C_+&D\\
 D&C_-
 \end{array}
 \right)\Phi,\\
 C_+ 
	 &=& 
	 \sum_{k,i\omega}
	 \frac{(i\Omega-\eta_{k + q}-m_B)(i\omega-\eta_k + m_B)}
	 {\left(\left(i\Omega-\eta_{k + q}+m_A\right)
			\left(i\Omega-\eta_{k + q}-m_B\right)-\varepsilon_{k + q}^2\right)
	  \left(\left(i\omega-\eta_k-m_A\right)
		    \left(i\omega-\eta_k+m_B\right)-\varepsilon_k^2\right)},\\
 C_- 
	 &=& 
	 \sum_{k,i\omega}
	 \frac{(i\Omega-\eta_{k + q}+m_A)(i\omega-\eta_k-m_A)}
	 {\left(\left(i\Omega-\eta_{k + q}+m_A\right)
	 	\left(i\Omega-\eta_{k + q}-m_B\right)-\varepsilon_{k + q}^2\right)
	 	\left(\left(i\omega-\eta_k-m_A\right)
	 	\left(i\omega-\eta_k+m_B\right)-\varepsilon_k^2\right)},\\
 D &=& \sum_{k,i\omega}\frac{{{\varepsilon _{k + q}}{\varepsilon _k}}}
	 {\left(\left(i\Omega-\eta_{k + q}+m_A\right)
	 	\left(i\Omega-\eta_{k + q}-m_B\right)-\varepsilon_{k + q}^2\right)
	 	\left(\left(i\omega-\eta_k-m_A\right)
	 	\left(i\omega-\eta_k+m_B\right)-\varepsilon_k^2\right)}. 
\end{eqnarray}
\end{widetext}
In this paper, we do not discuss the lifetime of a spin wave in a ferrimagnet. 
We simply consider the spectral weight of its spin wave. 
\begin{eqnarray}
i\nu
	&=&
	\frac{1}{2}\left[
		\pm\left(\epsilon_1-\epsilon_2\right)
		+\sqrt{\left(\epsilon_1+\epsilon_2\right)^2-4\epsilon_3^2}\right],\\
\epsilon_1
	&=&
	z_1JS_B+z_2J_AS_A(1-\zeta_q),\\
\epsilon_2
	&=&
	z_1JS_A+z_2J_BS_B(1-\zeta_q),\\
\epsilon_3
	&=&
	z_1J\sqrt{S_AS_B}\gamma_q,\\
\zeta_q &\equiv& \frac{1}{z_2}\sum_{\rm e_i}e^{i{\bf q}{\bf e}_i}.
\end{eqnarray}
It is assumed that the number of second neighbor sites of A- and B-sites $z_2$ 
are the same. 
It is useful to see the $\nu$-integrated spectral weight $I(q)$ shown in 
Fig.~\ref{ferri}.
For $S_A/S_B=1$, i.e., an antiferromagnet, $I(q)$ linearly decreases to zero 
around 
(0,0). 
This fact means that the spin current cannot be generated in the AF 
insulator~\cite{ohnuma}, whereas for the ferrimagnet such as one with 
$S_A/S_B=4$, the spectral weight around (0,0) becomes finite. 
This is caused by the degeneracy of the dispersion relation of spin waves.
The spin waves in the AF insulator are degenerated, whereas those in the 
ferrimagnet are split with a 
magnitude of gap, $|S_A-S_B|$.
Usually, the spin current is generated by an ac magnetic field, a temperature 
gradient, and so on. 
These external fields excite the magnons around (0,0). 
However, if we could excite the magnons around ($\pi$,$\pi$) by some means, it 
is obvious that the generated spin current could be a few orders of magnitudes 
larger than 
the conventional one obtained using a ferromagnet. 
This would be one of the strong merits of antiferromagnetic spintronics.

\end{document}